\documentclass[reprint, amsmath, amssymb, aps, prb,]{revtex4-1}

\usepackage{graphicx}
\usepackage{dcolumn}
\usepackage{bm}

\usepackage{xcolor}

\begin{document}

\preprint{APS/123-QED}

\title{Photoinduced Screening Breakdown Mediated by Plasmon Excitations\\ 
        and Potential Instabilities in the System}

\author{Anatoley T. Zheleznyak}
\email{anatoley.zheleznyak@vesperix.com}
\author{Tom Wallace}
\affiliation{Vesperix Corporation, 803 West Broad Street, Suite 520, Falls Church, VA 22046 }

\date{\today}

\begin{abstract}

In this work, we estimate the effect of photoinduced screening breakdown of the Coulomb potential mediated by plasmons. In contrast to previous studies, we consider the contribution from the divergence of renormalized inverse electron permittivity, which enhances Coulomb interaction at longer ranges. Our work reveals that the Coulomb potential acquires oscillating terms, which for the charged plane case could lead to a spatial heterostructure. In the resonant case ($\omega_p = \Omega$), photoinduced Coulomb interaction becomes attractive for small values of momenta. This Coulomb interaction renormalization could potentially lead to instabilities in the electron system, such as superconductivity and/or charge density waves. We found possible divergence of the electron coupling constant and discuss herein a potential new mechanism of superconductivity based on electron-electron interaction renormalization via the plasmon oscillating electric field. We also briefly discuss possible photoinduced $\epsilon$ near zero conditions.
\begin{description}
\item[PACS numbers] 73.20 Mf, 78.20.-e, 74.90.+n, 74.10.+v
\end{description} 
\end{abstract}
\maketitle

\section{\label{sec:Intro} Introduction}
Photoinduced screening breakdown was first theoretically investigated in the early 1970s \cite{ref:SB_orig, ref:Shmelev_book}, and was later studied in more detail \cite{ref:SB_Shmelev_ATZ,ref:SB_ATZ} In this paper we study the implications of this theory caused by the divergence of the inverse electric permittivity, which has never been considered. Broadly, this phenomenon is based on the renormalization of Coulomb interaction in electron systems subjected to strong electromagnetic radiation (EMR). In particular, due to renormalization, the static electric permittivity acquires dependencies on the electric permittivity's high frequency component at the frequency of the EMR $(\Omega)$ and its harmonics:
\begin{equation}
\frac{1}{\epsilon_{eff}(\bold{q})} = \sum\limits_{n} \frac{J_n^2(\bold{aq})}{\epsilon(\bold{q}, n\Omega)}
\label{eq:eps_eff}, 
\end{equation}
where $\bold{a} = e\bold{F}/m\Omega^2$ is the amplitude of the electron oscillations, $J_n$ is the Bessel function of order $n$, and the EMR is considered in the dipole approximation 
$\bold{F}(t) = \bold{F} \sin(\Omega t$), $v \ll c$. Eq.~(\ref{eq:eps_eff}) has a generic nature and can be derived by calculating an electron density response to the external Coulomb field in first order perturbation theory. It can also be obtained in the mean-field or Random Phase Approximation (RPA) for a collisionless electron plasma ($\omega_p \tau \gg 1$, where $\omega_p^2 = ne^2/\epsilon_0m$ is the plasma frequency and $\tau$ is the relaxation time). The EMR is assumed to propagate through the media without causing inter-band transitions, which requires that its photon energy is less than the energy gap ($\Omega < E_g, \hbar = 1$).

The Coulomb field subjected to a renormalized electric permittivity (\ref{eq:eps_eff}) acquires photoinduced terms that manifest non-exponentially decaying spatial dependencies, representing screening breakdown. 
The physical interpretation of the photoinduced screening breakdown is straightforward: in the presence of EMR, the Coulomb field acquires time-dependent components at the harmonics of the EMR, which in turn are dependent on the high frequency component of the electric permittivity. These harmonics spread through the media as high frequency components without being screened, and are averaged to a stationary field by the oscillating electrons' field. Particularly,  it was shown in Refs \cite{ref:SB_orig, ref:Shmelev_book, ref:SB_Shmelev_ATZ, ref:SB_ATZ} that the static Coulomb potential acquires a photoinduced long-range component, which decays as a power law $1/r^3$. The effect of screening breakdown in essence occurs due to non-screened propagation of the high frequency Coulomb potential components. 

To our knowledge, Eq. (\ref{eq:eps_eff}) has only been applied to the study of Coulomb screening of a non-degenerate electron gas when $\epsilon(\bold{q}, \omega)$ was approximated as $1$, which led to a Coulomb potential of $\varphi(r) \sim F^2/r^3$ (Refs. \cite{ref:SB_orig, ref:Shmelev_book}), and to a degenerate electron gas, where, in the presence of EMR, the Coulomb potential acquires some additional Friedel oscillation harmonics \cite{ref:SB_Shmelev_ATZ}. 

What we believe to be the most interesting phenomenon --- that the photoinduced component of the inverse effective permittivity given by Eq. (\ref{eq:eps_eff}) diverges at $\epsilon(\bold{q}, n\Omega) = 0$ --- has never been studied. The physics of this divergence are simple: it occurs whenever internal plasmons with frequencies matching the EMR's harmonics spread the respective harmonics of the Coulomb potential without screening. This spread potential is ultimately averaged by the oscillating conducting electrons to yield a stationary component. This may be considered a manifestation of the optical-plasmon resonance, although it is not classical resonant absorption by any means since the EMR propagates through the media not causing intra or inter band transitions. 

\section{\label{sec:SB} Screening Breakdown}

Breakdown of Coulomb screening is the only phenomenon that has been studied as a direct result of Eq.~ (\ref{eq:eps_eff}), and, as previously noted, the divergence of the photoinduced component has not been considered in previous relevant works \cite{ref:SB_orig, ref:Shmelev_book, ref:SB_Shmelev_ATZ, ref:SB_ATZ}. As we show below, such photoinduced divergences of the static component of electric permittivity lead to significantly stronger long-range dependence of the screened Coulomb potential. Similar to the aforementioned publications \cite{ref:SB_orig, ref:Shmelev_book, ref:SB_Shmelev_ATZ, ref:SB_ATZ}, we consider the screening of a point charge and a charged plane. In the linear-in-intensity approximation ($a^2$), the photoinduced term of the static Coulomb potential has the form:
\begin{equation}
\varphi_{PI}(\bold{r}) = \int \frac{d\bold{q}}{(2\pi)^3} \frac{\rho({\bold q}) (\bold{aq})^2 \exp(i\bold{qr})} {2\epsilon_0 q^2 \epsilon(\bold{q}, \Omega)}
\label{eq:SB_PI}, 
\end{equation}
where $\rho({\bold q})$ is a Fourier component of the stationary charge density. Considering contribution from a divergent term only, we approximate the permittivity in the vicinity of the point of divergence as:
\begin{equation}
\frac{1}{\epsilon(q,\Omega)} \approx \frac{1}{A(1-q/q_0)}
\label{eq:eps_div}, 
\end{equation}
 where $A$ is the slope and $q_0$ is the zero crossing point for the frequency-dependent electric permittivity, which should be calculated numerically in most cases. 
 
\begin{figure}
\includegraphics{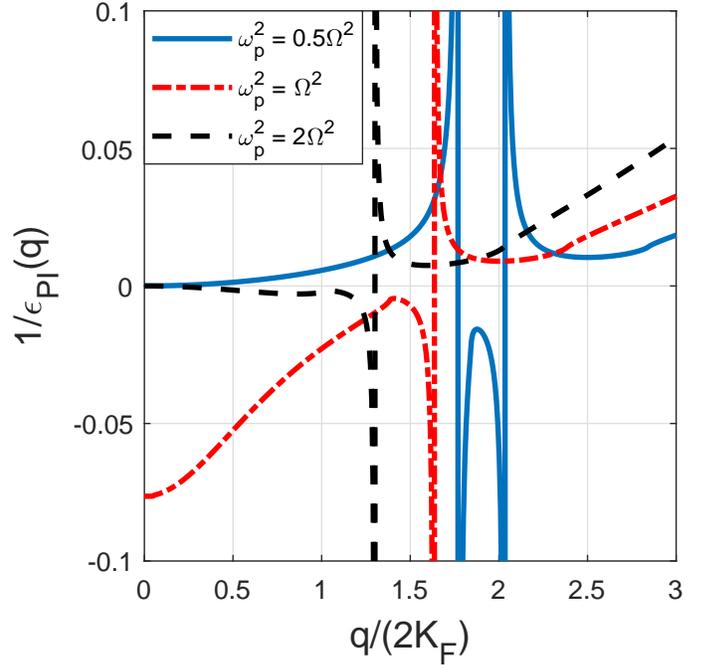}
\caption{\label{fig:EpsPI} Photoinduced divergence of the inverse static permittivity for illuminating frequencies near the plasma frequency.}
\end{figure}

 For the Lindhard form of the permittivity function, which corresponds to degenerate electron gas at zero temperature, there could be as many as two crossings of $\epsilon(q, \Omega) = 0$ at $\Omega \le \omega_p$ and one crossing at $\Omega > \omega_p$ as shown in Figure \ref{fig:EpsPI}. The zero-crossing of the frequency-dependent electric permittivity is a universal phenomenon; at $q = 0$ and $\Omega < \omega_p$, $\epsilon(q = 0, \Omega) < 0$, while for large values of $q$, $\epsilon(q \to \infty, \Omega) = 1$. There should therefore be at least a single zero-crossing for $\epsilon(q, \Omega)$ at $\Omega < \omega_p$. Retaining only the strongest long-range component of (\ref{eq:SB_PI}) for the point charge, ($\rho^{(PC)}(q) = Q$), Eq. (\ref{eq:SB_PI}) takes the form:
\begin{equation}
\varphi_{PI}^{(PC)}(\bold{r}) = \sum\limits_{i} \frac{Q(a \cos(\beta) q_0^{(i)})^2}{4\pi\epsilon_0 A_i} \frac{\cos(q_0^{(i)} r)}{r}
\label{eq:SB_PI_PC}, 
\end{equation}
where the summation $i$ is over the singularities and $\beta$ is the angle between $\bold{r}$ and $\bold F$. Similarly, for the charged plane ($\rho^{(CP)}(q) = 2\pi\sigma \delta(q_x)\delta(q_y)$ where $\sigma$ is the surface charge density), we get
\begin{equation}
\varphi_{(PI)}^{(CP)}(z) = \sum\limits_{i} \frac{\sigma a_z^2 q_0^{(i)})}{4\epsilon_0 A_i} \sin(q_0^{(i)} z)
\label{eq:SB_PI_CP}, 
\end{equation}
where $a_z$ is a projection of $\bold{a}$ vector to the $z$ axis.

Eqs. (\ref{eq:SB_PI_PC}, \ref{eq:SB_PI_CP}) show an essential enhancement of the potential caused by the singularity of the inverse static permittivity. The Coulomb potential of both a point charge (\ref{eq:SB_PI_PC}) and a charged plane (\ref{eq:SB_PI_CP}) acquire a non-screened structured component ($1/r$ for the point charge and constant for the charged plane), which oscillates with period $2\pi/q_0^{(i)}$. The periodic non-decaying structure of the charged plane potential (\ref{eq:SB_PI_CP}) points to the possibility of creating an artificial (photoinduced) superlattice/heterostructure similar to the structure of charge density waves. 

\subsection{Screening breakdown at resonance}

It is interesting to examine the resonance case $\Omega = \omega_p$. For a degenerate electron gas with parabolic dispersion $E_p = p^2/2m$, where $m$ is the effective mass, 
the Lindhard function for small $\bold{q}$ (up to quadratic terms) takes the form:
\begin{equation}
\epsilon(q, \Omega) = 1 - \frac{\omega_p^2}{\Omega^2} \big[1 + \frac{3(qk_F)^2}{5(m\Omega)^2}\big]
\label{eq:eps_Resonance},
\end{equation}
where $k_F$ is the Fermi momentum. As follows from (\ref{eq:eps_Resonance}), at $\Omega = \omega_p$, the inverse static electric permittivity (\ref{eq:eps_Resonance}) becomes a negative constant 
for q = 0 with the only surviving component being the photoinduced term proportional to $a^2$:
\begin{equation}
\frac{1}{\epsilon_{eff}(\bold{q} = 0)} = - \frac{5(a_qm\Omega)^2}{3k_F^2}
\label{eq:eps_eff_q=0}, 
\end{equation}
where $a_q$ is the projection of $\bold{a}$ in the direction of $\bold{q}$. Previously, we considered only terms linear in the intensity of the radiation ($a^2 $). $1/\epsilon_{eff}(q = 0)$ (\ref{eq:eps_eff_q=0}) is also linear in intensity. However, for $q=0$ at resonance, only terms linear in inetnsity survive, so Eq. (\ref{eq:eps_eff_q=0}) is valid for arbitrary values of intensity. Even though Eq. (\ref{eq:eps_eff_q=0}) is derived for the degenerate electron gas where electric permittivity is described by the Lindhard function, a negative sign of $1/\epsilon_{eff}(q = 0)$ may still be found, as long as $\epsilon(q, \Omega)$ is decreasing for small $q$, resulting in $1/\epsilon_{eff}(q = 0)< 0$. In the absence of radiation, $1/\epsilon(q = 0) \to 0$, which means that small values of $q$ do not contribute to the Coulomb potential. We believe that we may have encountered an universal phenomenon, the renormalization by the EMR of the Coulomb interaction becomes attractive at resonance conditions, which might have many fundamental consequences.  The photoinduced component of the Coulomb potential that comes from the small values of $q$ has the form:
\begin{eqnarray}
\varphi_{PI}^{(PC)}(\bold{r}) =&& -[2(2\pi - 1) \cos^2(\beta) + \sin^2(\beta)] \nonumber\\
 &&\times\frac{5Q(am\Omega)^2}{96\pi^2\epsilon_0 k_F^2}\frac{1}{r} 
\label{eq:SB_PI_PC_Resonance}, \\
\varphi_{PI}^{(CP)}(z) = && \frac{5}{12}\frac{\sigma (a_zm\Omega)^2}{\epsilon_0 k_F^2} |z|
\label{eq:SB_PI_CP_Resonance}. 
\end{eqnarray}
Eqs. (\ref{eq:SB_PI_PC_Resonance}, \ref{eq:SB_PI_CP_Resonance}) show that at resonance, the Coulomb potential is not screened for either the point charge or the charged plane. Note that for the charged plane (\ref{eq:SB_PI_CP_Resonance}), the interaction's sign is the opposite of the non-screened potential ($\varphi_0^{(CP)}(z) = - \sigma|z|/(2\epsilon_0$)). It is essential to note that the interaction's sign is flipped --- repulsion is replaced by attraction, which could have fundamental consequences for instabilities in the system. 

\subsection{Photoinduced superconductivity}

One interesting possible instability is the formation of plasmon-mediated electron pairs, similar to phonon-mediated Cooper pairs, resulting in a new mechanism for photoinduced superconductivity. To our knowledge, this is the first examination of possible Coulomb attraction of non-opposite charged particles via the electron-plasmon interaction in the framework of RPA. (Previously, the attractive component of the Coulomb potential for electron-electron interaction was only found beyond the framework of RPA \cite{ref:Uspenskii_Uspehi} due to exchange and correlation effects.)

It is well-known that logarithmic or weak divergences in susceptibility lead to various types of instabilities, including superconductivity, charge, and spin density waves. Logarithmic divergence of derivative of the inverse permittivity  lead to Friedel oscillations of the Coulomb potential. Eq. (\ref{eq:eps_eff}) manifests a linear divergence of the inverse electric permittivity as well as non vanishing contribution for $q = 0$ at resonance conditions, and the impact of such singularities has never been studied. Since logarithmic divergences are considered to be weak and nevertheless they lead to many fundamental instabilities in the system, the potentially stronger divergence of the inverse electric permittivity may also lead to instabilities in the system.  
As we mentioned above, it is possible that the photoinduced Coulomb attraction manifested by Eqs. (\ref{eq:SB_PI_PC}, \ref{eq:SB_PI_PC_Resonance}) could lead to a plasmon-mediated superconducting state. The theoretical framework for plasmon mediated superconductivity has been extensively studied, see for example Refs. \cite{ref:Kirzhnits_SC_DRF,ref:GKbook}, where the authors not only anticipated the discovery of high temperature superconductivity, but tried to find its plausible scenario. Specifically, the authors found that superconductivity could occur if Coulomb repulsion was suppressed, while high frequency oscillations in the electron system could lead to an increase in critical temperature ($T_c$). 

Photoinduced screening breakdown may provide these two fundamental conditions for the  superconducting state. In the vicinity of the singular points and for $q=0$ in the resonance case (\ref{eq:SB_PI_CP_Resonance}), $1/\epsilon_{eff}(\bf{q})$ becomes negative, which leads directly to Coulomb attraction. Thus, the screening breakdown has the potential to have a much stronger impact, not in suppressing Coulomb repulsion, but leading to direct attraction for some values of $q$. The presence of EMR also creates an additional oscillating mode in the electron system. (Note that plasmon superconductivity has been previously advanced as a mechanism for high-$T_c$ material \cite{ref:Pashitskii}; although the evidence for this is unclear. Nevertheless, as mentioned in \cite{ref:mazin_VL} the topic of plasmon superconductivity still keep coming regularly.) 
 
We can estimate the characteristics of this photoinduced superconducting state using the following analysis.
The plasmon coupling constant, $\lambda_{pl}$ is expressed in terms of the electric permittivity as \cite{ref:GKbook,ref:Pashitskii}:
\begin{equation}
\lambda_{pl} = -\alpha \int_{0}^{2 k_F} \frac{dq}{q} \frac{1}{\epsilon_{eff}(q)}
\label{eq:lambda_pl}.
\end{equation} 
At resonance at $q \to 0$ and in the vicinity of the singular points ($q_0$), where $1/\epsilon_{eff}(q) \to - \infty$ , the integral (\ref{eq:lambda_pl}) diverges logarithmically. We shall see that this divergence may lead to a higher critical temperature.

The largest currently known value of the electron-phonon coupling constant is in the low single digits; for example, it is somewhat more than $2$ for Pb-Bi alloys \cite{ref:mazin_VL}. The singularity of integral (\ref{eq:lambda_pl}) could potentially increase $\lambda_{pl}$ above this value, which in turn could lead to increase of the critical temperature $T_c$ to values similar to the plasmon energy $\hbar\omega_p$ or the Fermi energy ($E_F$). The same conclusion can also be reached in the framework of the standard BCS theory, using the attractive renormalized Coulomb interaction. The equation for the superconducting gap $\Delta_{\bold p}$ can be written as \cite{ref:Abrikosov_book}:
\begin{equation}
\Delta_{\bold p} = - \frac{1}{2\epsilon_0} \sum_{\bold p'} \frac{\Delta_{\bold p'}}{ \epsilon_{eff}(\bold p - \bold p') } \frac{4\pi e^2}{(p-p')^2}\frac{1 - n_{\bold p', +} - n_{-\bold p', -}}{[E_{\bold p'}^2 + \Delta_{\bold p'}^2]^{1/2}} 
\label{eq:gap_pl},
\end{equation} 
where $n_{\bold p, \pm}$ is the Fermi distribution function with momenta $\bold p$ and spin $\pm$. Due to the divergence of equation (\ref{eq:gap_pl}) at the singular points $q_0 = p - p'$, and at $p - p' = 0$ for the resonance case, this equation's standard BCS solution yields $T_c \sim \Delta E$, where $\Delta E$ is the energy scale in the electron system, $\Delta E \sim {E_F, \omega_p, \Omega}$, which is consistent with large values of $\lambda_{pl}$ as follows from Eq. (\ref{eq:lambda_pl}). In the long-term quest for high temperature superconductivity, the small size of $\lambda$ was considered a fundamental obstacle \cite{ref:Abrikosov_book}; screening breakdown offers a potential way to avoid this.
 
On the other side of the singularities, where $1/\epsilon_{eff}(q) \to+ \infty$, Coulomb repulsion is enhanced. As found in Ref. \cite{ref:Kirzhnits_SDF}, under condition 
$0 < \epsilon_{eff}(q,0) < 1$, the electron system supports charge density wave instabilities, which in our case occurs at $1/\epsilon_{eff}(q) \to+ \infty$. To some degree, this situation is similar to that seen in high-$T_c$ materials, the photoinduced renormalization of the Coulomb interaction could lead to the interplay of superconducting and charge density wave instabilities. It is difficult to predict which of the competing instabilities would dominate. The superconducting and charge density wave instabilities might neutralize each other for $q_0 \ne 0$, but the resonance $q =0$ instability would remain dominant \footnote{As the U.S. Naval Research Laboratory's Dr. Igor Mazin informed the authors, and as discussed in Ref. \cite{ref:mazin_VL},  screening breakdown should also affect phonon modes in a similar manner, which may lead to crystal instabilities.}. 

\subsection{Photoinduced $\epsilon$ near zero conditions}

Divergence of the inverse electric permittivity (\ref{eq:eps_eff}) is also seen in metamaterials, which are specifically characterized by $\epsilon$ taking values which are near zero or negative (for a review see Ref. \cite{Ref:Kinsler_MM}). In most cases, metamaterials require fabrication of a structure with resonant properties, and the ENZ properties are only obtained for a certain range of frequencies.

In our case, however, ENZ conditions are obtained for bulk materials (i.e., semiconductors) through the divergence of inverse electric permittivity (\ref{eq:eps_eff}) and/or in the resonance case for $q = 0$, and the metamaterial properties are controlled externally by the presence of EMR. The divergence of static electric permittivity also opens up the possibility of having metamaterials with ENZ over a wide range of frequencies, as long as stationary conditions are valid $\omega \ll (\omega_p, \Omega)$. 

Similar to hyperbolic metamaterials \cite{Ref:Smolyaninov_book}, photoinduced ENZ conditions also have spatial dependence due to EMR polarization. In particular, ENZ conditions should require a specific photon momentum $k = \omega/c$. For EMR in the infrared frequency range, the values of $k$ are very small compared to the characteristic values in the system, such as Thomas-Fermi screening length and Fermi momentum $k_F$, so divergence should only be considered for small values of $q$ in close proximity to the resonance case ($\Omega \approx \omega_p$) (\ref{eq:eps_Resonance}). A solution of $\epsilon(k, \Omega) = 0$ produces wave vector $k_0$, which yields the divergence of the inverse static permittivity:
\begin{equation}
k_0^2 = \frac{5(m\Omega)^2}{3k_F^2} \big( 1 - \frac{\omega_p^2}{\Omega^2} \big). 
\label{eq:k_0} 
\end{equation}
The range of wave vector $\delta k$ where photoinduced inverse dielectric permittivity is larger than some threshold value $B$ ($B \gg 1$) can be found as:
\begin{equation}
\frac{\delta k}{k_0} = \frac{5(a m \Omega)^2}{12 B k_F^2}, 
\label{eq:dk_over_k_0} 
\end{equation}
where $\delta k$ is a small value, since it is proportional to $a^2$, where $a$ is the amplitude of the electron oscillations, which are typically the smallest length scale in the system, divided by the large factor $B \gg 1$. Thus, for the stationary case, ENZ is caused by external radiation with frequency $\Omega$. Note that the ENZ frequency range is still restricted by the variation of wave vector $\delta k$ in the vicinity of $k_0$ (\ref{eq:k_0}--\ref{eq:dk_over_k_0}). 

Of particular interest is a non-stationary case, where the renormalized inverse electric permittivity takes the form:
\begin{equation}
\frac{1}{\epsilon_{eff}(\omega,\bold{q})} = \sum\limits_{n} \frac{J_n^2(\bold{aq})}{\epsilon(\bold{q}, \omega + n\Omega)}
\label{eq:eps_eff_w}. 
\end{equation}
Similar to the stationary case, we may consider the divergence of the photoinduced component of (\ref{eq:eps_eff_w}) at the frequency of EMR ($\Omega$):
\begin{equation}
\frac{1}{\epsilon^{(PI)}_{eff}(\bold{q}, \Omega)} = \frac{(\bold{aq})^2}{4}\Big[{\frac{1}{\epsilon(\bold{q}, 2\Omega) }+ \frac{1}{\epsilon(\bold{q})}}\Big]
\label{eq:eps_PI_w}. 
\end{equation}
The divergence of (\ref{eq:eps_PI_w}) at $2\Omega \approx \omega_p$ occurs at EMR wave vector $k_0$:
\begin{equation}
k_0^2 = \frac{20(m\omega)^2}{3k_F^2} \big( 1 - \frac{\omega_p^2}{4\Omega^2} \big) 
\label{eq:k_0_w}.
\end{equation}
The factor $1-\omega_p^2/4\Omega^2$ can be adjusted such that $k_0 \approx \Omega/c$ and the range of the wave vector where $|(aq)^2/4\epsilon(q,2\Omega)|$ is above the threshold $B$ is:
\begin{equation}
\frac{\delta k}{k_0} = \frac{5(a m \omega)^2}{6 B k_F^2}. 
\label{eq:dk_over_k_0_w}
\end{equation}
Thus, if the EMR is not strong, the non-linear (photoinduced) terms are not relevant; a very narrow frequency region of  ENZ is produced, which may not include the radiation frequency $\Omega$. Increasing the EMR's intensity would also increase the frequency range (\ref{eq:dk_over_k_0_w}), causing the dielectric permittivity to eventually become divergent at the EMR's frequency, which would drastically modify such material optical properties as reflectivity. 

Overall, the analysis of the photoinduced screening breakdown points to many conceptual similarities with the phenomena observed in metamaterials. Firstly, the ENZ conditions for metamaterials have spatial dependence, which comes from a special fabrication of the metamaterials from a mixture of metals and dielectrics. In our case, the ENZ is also spatially dependent due the EMR polarization, particularly at the momenta $\bf{q} \perp \bf{a}$ all photoinduced effects vanish. Secondly, it was found that the superconducting critical temperature is tripled for the Aluminum metamaterials as compared to a pure Aluminum\cite{Ref:Smolyaninov_ESC} and the effect is attributed to a renormalization of the dielectric permittivity due to a mixture of Aluminum with ferroelectric nanoparticles. This similarity with our case is just conceptual since we only point to a possibility of having the photoinduced superconductivity orchestrated by the screening breakdown. Lastly, Dr. Igor Smolyaninov came up with a scheme for the oscillating field pattern by the charge point source \cite{Ref:Kinsler_MM}, which at this point is only confirmed by simulations. Even though the oscillating nature of the point charge field have not been elaborated in any details, it is quite possible that it comes from the divergence of the inverse dielectric permittivity, similar to the Coulomb screening breakdown that we have discussed in this paper. Our major conceptual difference with metamaterials is that the phenomena described in this paper should be observable in regular materials under the influence of the EMR.   

\subsection{Estimates}

Photoinduced screening breakdown derivations are obtained for a generic situation of free electron gas being subjected to strong radiation, where only a few criteria need to be satisfied, one being no interband transition ($\Omega < E_g$). 
Since the effect is proportional to the square of the amplitude of the electron oscillations ($a^2$), favorable conditions would occur in a material with a small effective electron mass and a relatively low EMR frequency. Estimates are made for a semiconductor with small effective mass, n-doped $InAs$, and EMR in the far infrared wavelength at $ \lambda = 20 \mu m$. 

The concentration of carriers that corresponds to condition $\omega_p = \Omega$ is $n = 6.41\times10^{16} cm^{-3}$. Choosing the value of the electromagnetic field $F = 10^6 V/m$, the power density of the EMR is $0.26 MW/cm^2$. For these parameters, the divergence of the photoinduced inverse static permittivity is illustrated in Fig. \ref{fig:EpsPI}. Choosing the range $r = 1.9\times10^{-7} cm$, which is equal to five inverse Thomas-Fermi screening lengths, we can estimate the Coulomb potentials from Eqs. (\ref{eq:SB_PI_PC}) and (\ref{eq:SB_PI_PC_Resonance}). For Eq. (\ref{eq:SB_PI_PC}) (non-resonance case) estimating the slope of electric permittivity $A_i = 1$ and a zero crossing point of $q_0^{(i)}=2k_F$, the photoinduced Coulomb potential of the point charge is $\varphi_{PI}^{(PC)} (r)=13.74 meV$, while for Eq. (\ref{eq:SB_PI_PC_Resonance}) (the resonance case) the potential of the point charge is $\varphi_{PI}^{(PC)} (r)= - 1.34 meV$. 

For the same parameters, the values of $k_0 = 1.95\times10^5 m^{-1}$ (\ref{eq:k_0}) for $1- w_p^2/\Omega^2 = 10^{-6}$, which is in the infrared frequency range, and the ratio (\ref{eq:dk_over_k_0}) 
$\delta k /k_0 = 0.07\% $ at $B = 10$. Increasing the EMR intensity by 2 orders of magnitude would expand $\delta k /k_0$ to $7\% $, which would transform the material to non-reflective in a wide frequency range. 

\subsection{Could screening breakdown lead to superconductivity in the absence of EMR?}

Finally, we discuss a more speculative scenario. Plasmon waves are, in essence, electron density fluctuations, which create an AC electric field at the plasma frequency. These electron density fluctuations to some degree are analogous to the external field. Like external fields, plasmon waves create high frequency harmonics of the Coulomb potential, which are spread in a similar manner by internal plasmons without screening, and are also averaged to a stationary component by the AC field in a self-consistent manner. In other words, it may be possible for Coulomb screening breakdown to occur and persist in the absence of external radiation. In this case, the plasmons' electron density oscillations would cause a similar divergence of inverse static electric permittivity and lead to all of the effects previously discussed in this paper. For $q = 0$, we may obtain electron-electron attraction and the formation of Cooper-type pairs, which would lead to superconductivity via plasmon-mediated Coulomb interaction. 

To our knowledge, this potential superconductivity mechanism has never been discussed. Simple estimates show that electron plasma density fluctuations in the frequency range of $\omega_p$ create an internal AC field $F \sim \sqrt{n E_F/\epsilon_0}$, which is in the range $F \sim 3\times10^6$ (V/m) for the example parameters in the previous section. Note that for the Maxwell distribution function, the amplitude of the internal AC electric field is defined by the energy scale $F \sim \sqrt{n T/\epsilon_0}$ (see Ref. \cite{Ref:Krall_Plasma_Book}). Obviously, these fluctuations are not coherent; however, there could be a range where the coherence length is large enough for Cooper-type pair formation. Note that in cuprate superconduuctors, the size of the Cooper pairs can  be as small as $3\times10^{-9}$ m, while the wavelength of the plasmon is on the order of $\lambda_p \sim 10^{-6}$ m) ($\omega_p \sim 1 eV$ is a typical value of the plasmon energy in cuprates.) As we mentioned in our estimates, the screening breakdown effect becomes larger when the effective mass is smaller, which may be particularly relevant to high-$T_c$ materials that have a near flat Fermi surface. 

\section{\label{sec:Intro} Conclusion}

We have examined the effect of photoinduced Coulomb screening breakdown mediated by plasmons. This phenomenon originates from the divergence of renormalized inverse electron permittivity, and leads to the appearance of long-range Coulomb interaction with non-screened  behavior. We considered screening breakdown for two idealized cases, the point charge and the charged plane. Our studies showed that due to divergence of inverse effective electric permittivity, the Coulomb potential acquires oscillating terms, which for the charged plane could lead to a spatial heterostructure. In the resonance case ($\omega_p = \Omega$), photoinduced Coulomb interaction becomes attractive. This renormalization of the Coulomb interaction could also lead to potential instabilities in the electron system, such as superconductivity and/or charge density waves. We also illustrated a potential divergence of the coupling constant and discussed the possibility of a superconducting state. Moreover, we speculated that electron-electron attraction could take place via renormalization of the electric permittivity by plasmon electric field. Photoinduced ENZ conditions were also considered. 

\begin{acknowledgments}
The authors are grateful to Drs. Igor Smolyaninov, Igor Mazin, Dr. Mike Osofsky and Prof. Victor Yakovenko for useful discussions. Anatoley Zheleznyak also expresses gratitude to Prof. G. M. Shmelev for initial seed ideas. 
\end{acknowledgments}

\end{document}